\documentclass[aps,prd,onecolumn,groupedaddress,showpacs,nofootinbib,amssymb]{revtex4}
\usepackage{graphicx,bm,color}
\usepackage{amsmath}
\usepackage{amssymb}
\usepackage{amsfonts}

\newcommand{\be}{\begin{equation}}
\newcommand{\ee}{\end{equation}}
\newcommand{\bea}{\begin{eqnarray}}
\newcommand{\eea}{\end{eqnarray}}
\newcommand{\beaa}{\begin{eqnarray*}}
\newcommand{\eeaa}{\end{eqnarray*}}

\newcommand{\nn}{\nonumber \\}
\newcommand{\e}{\mathrm{e}}

\allowdisplaybreaks[4] 

\begin{document}

\title{Stable phantom-divide crossing in two scalar models 
with matter}

\author{Rio Saitou$^1$ and Shin'ichi Nojiri$^{1,2}$}

\affiliation{
$^1$ Department of Physics, Nagoya University, Nagoya
464-8602, Japan \\
$^2$ Kobayashi-Maskawa Institute for the Origin of Particles and
the Universe, Nagoya University, Nagoya 464-8602, Japan}

\begin{abstract}

We construct cosmological models with two scalar fields, 
which has the structure as in the ghost condensation model 
or $k$-essence model. 
The models can describe the stable phantom crossing, which should be 
contrasted with one scalar tensor models, where the infinite instability 
occurs at the crossing the phantom divide. 
We give a general formulation of the reconstruction in terms of the e-foldings $N$ 
by including the matter although in the previous two scalar models, 
which are extensions of the scalar tensor model, 
it was difficult to give a formulation of the reconstruction when we include 
matters. 
In the formulation of the reconstruction, we start with a model with some 
arbitrary functions, and find the functions which generates the 
history in the expansion of the universe. 
We also give general arguments for the stabilities of the models and the 
reconstructed solution. 
The viability of a model is also investigated by 
comparing the observational data. 

\end{abstract}

\pacs{95.36.+x, 98.80.Cq}

\maketitle

\section{Introduction \label{SecI}}

The observation of the cosmic microwave background radiation (CMBR) shows that
the present universe is spatially flat 
\cite{de Bernardis:2000gy,Hanany:2000qf}.
Hence, the starting point  for current cosmology is
spatially flat Friedman-Lemaitre-Robertson-Walker (FLRW) universe,
whose metric is given by
\be
\label{JGRG14}
ds^2 = - dt^2 + a(t)^2 \sum_{i=1,2,3} \left(dx^i\right)^2\, .
\ee
Here $a$ is called  the scale factor. 
The FLRW equations in the Einstein gravity coupled with perfect 
fluid are well-known to be:
\be
\label{JGRG11}
\rho =\frac{3}{\kappa^2}H^2 \, ,\quad
p = - \frac{1}{\kappa^2}\left(3H^2 + 2\dot H\right)\, .
\ee
Here the Hubble rate $H$ is defined by $H\equiv \dot a/a$ and dot denotes the derivative with
respect to the cosmic time $t$.

By the observation of the type Ia supernovae, we now believe that the 
expansion of the present universe is
accelerating \cite{Perlmutter:1998np,Perlmutter:1997zf,Riess:1998cb}.
What generates the accelerating expansion of the universe is called dark
energy (for recent review, see 
\cite{Silvestri:2009hh,Li:2011sd,Caldwell:2009ix}). 
The dark energy could be a perfect fluid with negative equation of state 
(EoS) parameter . 
The future of the universe is mainly governed by the equation of state 
parameter $w_\mathrm{DE}$ of the dark energy 
$w_\mathrm{DE}\equiv p_\mathrm{DE} / \rho_\mathrm{DE}$,
where $p_\mathrm{DE}$ and $\rho_\mathrm{DE}$ are the pressure and the energy 
density of the dark energy.
When $w_\mathrm{DE} <-1/3$, the accelerating expansion can be generated 
and the observational data indicates that $w_\mathrm{DE}$ is close to $-1$.
If the $w_\mathrm{DE}$ is exactly $-1$, the present universe is described by 
the $\Lambda$CDM model, where the accelerating expansion is generated by 
the cosmological term 
and the universe evolves into the asymptotic de Sitter space-time.
The dark energy with $-1<w_\mathrm{DE}<-1/3$ is called as quintessence and that
with $w_\mathrm{DE}<-1$ as phantom \cite{Caldwell:1999ew}. 
If the dark energy is the phantom, 
the future universe usually evolves to a
finite-time future singularity called  Big Rip \cite{astro-ph/0302506}, where
the scale factor $a$ of the universe will diverge in the finite future 
(for other type of singularities, see \cite{Barrow:2004xh,Barrow:2004hk}
and for the classification of the singularities, see \cite{hep-th/0501025}).

Recent cosmological data seems to indicate that there occurred the crossing of
the phantom divide line in the near past (see, for example, 
\cite{Nesseris:2006er}),
that is, the equation of state (EoS) parameter $w_\mathrm{DE}$ crossed 
the line $w_\mathrm{DE}=-1$.

In this paper, we consider the models with two scalar fields to describe the 
phantom crossing. 
It is known that in one scalar tensor models, the large instability occurs when
crossing the phantom divide (or cosmological constant border) \cite{Vikman:2004dc}. 
We now develop phantom cosmology described by two-scalar tensor theory 
which represents a kind of quintom model
\cite{arXiv:astro-ph/0404224,hep-th/0405034,Ito:2011ae}
(for review, see \cite{arXiv:0909.2776} and for generalizations,
see \cite{arXiv:0811.3643, Nojiri:2005pu}).
In case of two scalar tensor, we can construct a model which is stable at the 
phantom crossing. In the previous work, it has not been succeeded to a general 
formulation of the reconstruction of the two scalar tensor model when we include 
{\it matters}. 
In the formulation of the reconstruction, we start with a model with some 
arbitrary functions, and find the functions which generates the 
history in the expansion of the universe, which complies with
observational data (for the reconstruction in $F(R)$ gravity, see 
\cite{arXiv:0908.1269} and the reconstruction for the general models, 
see \cite{arXiv:1011.0544}). 
In this paper, in terms of the e-foldings $N$, 
we consider the two scalar models whose structures are 
similar to the $k$-essence models 
\cite{Chiba:1999ka,ArmendarizPicon:2000dh,ArmendarizPicon:2000ah}
or ghost condensation models 
\cite{ArkaniHamed:2003uy,ArkaniHamed:2003uz}. 
In the models, it is straightforward to include matters and 
it makes us to be able to consider the models with the phantom crossing 
and the matter dominant era. 

\section{Cosmological reconstruction by one scalar model and stability \label{SecII}}

Before going to the two scalar model, we clarify the problem in the one scalar model, 
that is, there appears infinite instability at the phantom crossing \cite{Vikman:2004dc}. 
This section is based on \cite{Nojiri:2005pu,Capozziello:2005tf} and 
we start with the following action:
\be
\label{ma7}
S=\int d^4 x \sqrt{-g}\left\{
\frac{1}{2\kappa^2}R - \frac{1}{2}\omega(\phi)\partial_\mu \phi
\partial^\mu\phi - V(\phi) + \mathcal{L}_\mathrm{m} \right\}\, .
\ee
Here, $\omega(\phi)$ and $V(\phi)$ are functions of the scalar field $\phi$.
The function $\omega(\phi)$ is not relevant and can be absorbed into the
redefinition of the scalar field $\phi$ as follows,
\be
\label{ma13}
\varphi \equiv \int^\phi d\phi \sqrt{\left|\omega(\phi)\right|} \, .
\ee
Then the kinetic term of the scalar field in the action (\ref{ma7}) has the
following form:
\be
\label{ma13b}
 - \omega(\phi) \partial_\mu \phi \partial^\mu\phi
= \left\{ \begin{array}{ll}
 - \partial_\mu \varphi \partial^\mu\varphi &
\mbox{when $\omega(\phi) > 0$} \\
\partial_\mu \varphi \partial^\mu\varphi & \mbox{when $\omega(\phi) < 0$}
\end{array} \right. \, .
\ee
The case of $\omega(\phi) > 0$ corresponds to the quintessence or
non-phantom scalar field, but the case of $\omega(\phi) < 0$ corresponds
to the phantom scalar.
Although $\omega(\phi)$ can be absorbed into the redefinition of the
scalar field, we keep $\omega(\phi)$ since the transition between the
quintessence and the phantom can be described by the change of the sign
of $\omega(\phi)$.

For the action (\ref{ma7}), the energy density and the pressure of the scalar 
field are given as follows:
\be
\label{ma8}
\rho = \frac{1}{2}\omega(\phi){\dot \phi}^2 + V(\phi)\, ,\quad
p = \frac{1}{2}\omega(\phi){\dot \phi}^2 - V(\phi)\, .
\ee
If the potential $V$ is positive, the EoS parameter $w$ is greater than $-1$
if $\omega(\phi)$ is positive but $w$ is less than $-1$
if $\omega(\phi)$ is negative.
Then the transition between the quintessence and the phantom can be 
described by the change of the sign of $\omega(\phi)$.
By using the FLRW equation (\ref{JGRG11}), the EoS parameter $w\equiv p/\rho$ 
can be expressed in the following form
\be
\label{i2}
w = - 1 - \frac{2\dot H}{3H^2}\, .
\ee
Then in the quintessence phase, where $1>1/3 w >-1$, we find $\dot H<0$ and
in the phantom phase, where $w<-1$, $\dot H>0$. Then on the point of the
transition between the quintessence and the phantom, $\dot H$ vanishes.

In order to consider and explain the cosmological reconstruction in terms of 
one scalar model, we rewrite the FLRW equation (\ref{JGRG11}) with the
expressions (\ref{ma8}) as follows:
\be
\label{ma9}
\omega(\phi) {\dot \phi}^2 = - \frac{2}{\kappa^2}\dot H\, ,\quad
V(\phi)=\frac{1}{\kappa^2}\left(3H^2 + \dot H\right)\, .
\ee
Assuming $\omega(\phi)$ and $V(\phi)$ are given by a single function
$f(\phi)$, as follows,
\be
\label{ma10}
\omega(\phi)=- \frac{2}{\kappa^2}f_{,\phi}(\phi )\, ,\quad
V(\phi)=\frac{1}{\kappa^2}\left(3f(\phi)^2 + f_{,\phi}(\phi)\right)\, ,
\ee
where $,\phi \equiv \frac{\partial}{\partial \phi}$, we find that 
the exact solution of the FLRW equations
(when we neglect the contribution from the matter)
has the following form:
\be
\label{ma11}
\phi=t\, ,\quad H=f(t)\, .
\ee
It can be confirmed that the equation given by the variation over $\phi$
\be
\label{ma12}
0=\omega(\phi)\ddot \phi + \frac{1}{2}\omega_{,\phi}(\phi){\dot\phi}^2
+ 3H\omega(\phi)\dot\phi + V_{,\phi}(\phi)\, ,
\ee
is also satisfied by the solution (\ref{ma11}).
Then, the arbitrary universe evolution expressed by $H=f(t)$ can be
realized by an appropriate choice of $\omega(\phi)$ and $V(\phi)$.
In other words, by defining the particular type of universe evolution,
the corresponding scalar-Einstein gravity may be found.

We now show that there occurs the very large instability when
crossing the cosmological constant line $w=-1$.
For this purpose, by introducing the new variables $X_\phi$ and $Y$
as follows
\be
\label{R1}
Z \equiv \dot \phi\, ,\quad Y \equiv \frac{f(\phi)}{H}\, ,
\ee
we rewrite the FLRW equation (\ref{JGRG11}) with (\ref{ma8})
and the field equation (\ref{ma12}) as 
\be
\label{R2}
\frac{d Z}{dN} = - \frac{f_{,\phi \phi}(\phi) 
\left( Z^2 - 1 \right)}{2 f_{,\phi}(\phi) H}
 - 3 \left( Z - Y \right) \, ,\quad
\frac{d Y}{dN} = \frac{f_{,\phi}(\phi) \left( 1 - Z Y \right) Z}{H^2}\, .
\ee
Here $N$ is called as e-foldings, 
defined by the scale factor $a$ as $a=\e^{N-N_0}$, where $N_0$ is the value of $N$ 
in the present universe. 
Then we find $d/dN\equiv H^{-1}d/dt$.
Since we have $Z=Y=1$ for the solution (\ref{ma11}), we now consider the 
following perturbation:
\be
\label{R3}
Z = 1 + \delta Z\, , \quad Y = 1 + \delta Y\,
\ee
Then
\be
\label{R4}
\frac{d}{dN} \left( \begin{array}{c}
\delta Z \\
\delta Y
\end{array} \right)
= \left( \begin{array}{cc}
 - \frac{\ddot H}{\dot H H} - 3 & -3 \\
 - \frac{\dot H}{H^2} & - \frac{\dot H}{H^2}
\end{array} \right) \left( \begin{array}{c}
\delta Z \\
\delta Y
\end{array} \right)\, .
\ee
Here, the solution (\ref{ma11}) is used.
The eigenvalues $M_\pm$ of the matrix are given by
\be
\label{R5}
M_\pm = \frac{1}{2} \left\{ - \left( \frac{\ddot H}{\dot H H}
+ \frac{\dot H}{H^2} + 3 \right) \pm \sqrt{ \left( \frac{\ddot H}{\dot H H}
+ \frac{\dot H}{H^2} + 3 \right)^2 - \frac{4\ddot H}{H^3}} \right\}\, .
\ee
In order that the solution (\ref{ma11}) could be stable, all the eigenvalues
$M_\pm$ must be negative. One now considers the region near the transition 
between the quintessence phase and the phantom phase, where $\dot H\sim 0$.
When we consider the transition from the quintessence phase, where $\dot H<0$,
to the phantom phase, we find $\ddot H > 0$.
On the other hand, when we consider the transition from the phantom phase,
where $\dot H>0$, to the quintessence phase, we find $\ddot H < 0$.
Then for both transitions, one finds $\ddot H / \dot H H$ is large
and negative. The eigenvalues $M_\pm$ are given by
\be
\label{R6}
M_+ \sim - \frac{\ddot H}{\dot H H}\, ,\quad M_- \sim 0\, .
\ee
Then $M_+$ is positive and diverges on the point of the transition $\dot H=0$.
Hence, the solution describing the transition is always unstable and the 
instability diverges at the transition point and therefore  the transition 
is prohibited in the one scalar model\footnote{
This does not always mean that any one scalar model with non-canonical kinetic 
term shows the instability (see \cite{Deffayet:2010qz}, for example).  
}.

\section{Cosmological reconstruction by two scalar model \label{SecIII}}

We now consider the new type of two scalar models and give a formulation of the 
cosmological constant. The two scalar models are different from those proposed 
in the previous papers \cite{arXiv:astro-ph/0404224,hep-th/0405034,Ito:2011ae} 
and the action with two scalar field $\phi$ and $\chi$ is given by 
\be
\label{two1}
S = \int d^4x \sqrt{-g}\left( \frac{1}{2\kappa^2} R +X + \omega(\phi)X^2
- U + \eta(\chi)U^2 - V(\phi , \chi) + \mathcal{L}_{m} \right)\, ,
\ee
and assume the flat FLRW metric (\ref{JGRG14}). 
Here $X$ and $U$ are defined by $X =-\frac{1}{2}\partial^\mu \phi \partial_\mu \phi$ 
and $U =-\frac{1}{2}\partial^\mu \chi \partial_\mu \chi $ and 
$\mathcal{L}_{m}$ is the Lagrangian density of the matters.   

We consider the models where there do not exist no direct interactions between 
the scalar fields and the ordinary matters. 
Then the conservation law of the matters is given by 
\be
\label{two2}
\rho_m'(N) + 3(\rho_m(N)+p_m(N)) = 0\, .
\ee
Here $\rho_m$ and $p_m$ are the energy density and the pressure of the matters, 
respectively and the prime $'$ denotes the derivative with respect to the e-folding number $N$. 

For the action (\ref{two1}), the FLRW equations have the following form: 
\bea
\label{two3}
X + \omega(\phi )X^2 - U + \eta (\chi )U^2 - V 
&=& -\frac{1}{\kappa ^2}\left( 2HH' +3H^2\right) + \rho_m +\frac{1}{3}\rho_m' \, ,\nn
X + 2\omega(\phi )X^2 - U + 2\eta (\chi )U^2 &=& -\frac{1}{\kappa ^2}HH' + \frac{1}{6}\rho_m'\, .
\eea
The energy density contributing to the Hubble rate can be divided into contributions from 
the scalar fields and the matter part as follows,
\be
\label{two4}
H^2(N) = \frac{\kappa ^2}{3}(\rho_s(N) + \rho_m(N))\, ,
\ee
and we rewrite the FLRW equations as follows,
\bea
\label{two5}
&& V(\phi,\, \chi) - \frac{1}{2}(X-U) = \rho_s(N) + \frac{1}{4}\rho_s'(N)\, , \\
\label{two5b}
&& \omega(\phi ) X^2 + \eta(\chi ) U^2 +\frac{1}{2}(X-U) = -\frac{1}{12}\rho_s'(N)\, .
\eea
The field equations for the scalar fields are given by
\bea
\label{two6}
V_{,\phi}(\phi,\, \chi) + 3\omega_{,\phi}(\phi )X^2 &=& - H^2(1+6\omega X)\phi'' 
- \left( HH'(1+6\omega X) + 3H^2 +6H^2\omega X\right)\phi' \, ,\\
\label{two7}
V_{,\chi}(\phi,\, \chi) + 3\eta_{,\chi}(\chi)U^2 &=& H^2(1-6\eta U)\chi'' 
+ \left( HH'(1-6\eta U) + 3H^2 -6H^2\eta U\right)\chi'\, ,
\eea
where $\phi \equiv \frac{\partial}{\partial \phi}$ and $\chi \equiv \frac{\partial}{\partial \chi}$ respectively.  
Then, if $V$, $\omega$, and $\eta $ satisfy the following relations
\bea
\label{two8}
&& V(mN,\, mN) = f(N) + \frac{1}{4}f'(N)\, , \nn
&& \omega(mN) + \eta (mN) = -\frac{f'(N)}{3m^4H^4(N)}\, , \nn
&& V_{,\phi }(mN,\, mN) + \frac{3}{4}m^4H^4(N)\omega_{,\phi }(mN) \nn
&& \quad = -m(3H^2(N) + HH'(N)) -3m^3H^2(N)\omega (mN)(H^2(N) + HH'(N))\, , \nn 
&& V_{,\chi }(mN,\, mN) + \frac{3}{4}m^4H^4(N)\eta_{,\chi }(mN) \nn
&& \quad = m(3H^2(N) + HH'(N)) -3m^3H^2(N)\eta (mN)(H^2(N) + HH'(N))\, ,
\eea
we can obtain the following solution
\be
\label{two9}
\phi =\chi = mN \, , \quad 
H^2(N) = \frac{\kappa^2}{3}(f(N) + \rho_m(N)) \, .
\ee

We now find explicit forms of the functions $\omega(\phi)$, $\eta(\chi)$, 
and $V(\phi,\,\chi)$ in the action (\ref{two1}), which satisfy the above relations.
We may choose, for example, the forms of $\omega(\phi)$ and $\eta(\chi )$ 
with two arbitrary functions $\alpha(\phi \, \mathrm{or}\,\chi)$ 
and $f(\phi/m\, \mathrm{or}\, \chi/m)$ as follows, 
\bea
\label{two10}
\omega(\phi ) &=& -\frac{1}{3m^3H^4(\phi/m)}\frac{\partial f(\phi /m)}{\partial \phi} 
+ \alpha(\phi ) \, , \\
\eta(\chi) &=& -\alpha(\chi) \, .
\eea
We also define a new function $\tilde f(\phi,\,\chi)$ by
\bea
\label{two11}
\tilde f(\phi, \chi) &=& 
 -m\left\{ \int d\phi \left(3m^2\omega(\phi )H^4(\phi/m)+2H^2(\phi/m)\right) \right. \nn
&& \left. + \int d\chi \left(3m^2\eta(\chi)H^4(\chi/m)-2H^2(\chi/m)\right) \right \}\, ,
\eea
which gives
\be
\label{two12}
\tilde f(mN,\,mN) = f(N)\, .
\ee
Using the function $\tilde f(\phi,\,\chi)$, we find $V(\phi, \, \chi)$ is given by
\bea
\label{two13}
V(\phi,\chi ) &=& - m\left( \int d\phi H^2(\phi/m)-\int d\chi H^2(\chi/m)\right) \nn
&&   + \tilde f(\phi, \chi) + \frac{m}{4}\left( \frac{\partial \tilde f}{\partial \phi}(\phi ) 
+ \frac{\partial \tilde f}{\partial \chi}(\chi ) \right)\, .
\eea
By reconstructing the action by using two functions, $\alpha(\phi \, \mathrm{or}\,\chi)$ and 
$f(\phi/m\,\mathrm{or}\,\chi/m)$, we can obtain the arbitrary history of the expansion 
of the universe as we desire.

\section{Stability \label{SecIV}}

In this section, we discuss the stability of the reconstructed solution by considering the perturbations. 
We also investigate the stability by using the sound speed. If the square of the sound speed is negative, 
the fluctuations grow up exponentially and the homogeneous universe becomes unstable.  

\subsection{Stability of the reconstructed solution}

In order to investigate the stability of the solution (\ref{two9}), 
we consider the fluctuation from the solutions. 
We define the 1st order perturbations from the solutions as follows,
\be
\label{two14}
\phi = \phi_0 + \delta \phi(N)\, ,\quad \chi =  \chi_0+ \delta \chi(N)\, ,\quad 
\dot \phi = \dot \phi_0 + \delta x(N)\, ,\quad \dot \chi = \dot \chi_0 + \delta y(N)\, ,
\ee
where $\delta x(N) = \delta \dot \phi(N)$ and $\delta y(N)=\delta \dot \chi(N)$.
We note that this system has four physical degrees of freedom (dof) and 
we identify $\phi$, $\chi$, $\dot \phi$, $\dot \chi$ with the dof here. 
It is also noted that the Hubble rate $H$ are not 
the dof because it can be removed by 
the constraint equation, i.e., the first FLRW equation (\ref{two4}). 
Substituting the expressions in (\ref{two14}) into Eqs.~(\ref{two5}), 
(\ref{two5b}), (\ref{two6}), and (\ref{two7}), 
we obtain the evolution equations of the 1st order fluctuations as follows,
\be
\label{two15}
\frac{d}{dN}
\begin{pmatrix}
\delta \phi \\
\delta \chi \\
\delta x \\
\delta y \\
\end{pmatrix}
= \begin{pmatrix}
0 &0 &H^{-1} &0 \\
0 &0 &0 &H^{-1} \\
a_{1} &a_{2} &a_{3} &a_{4} \\
a_{5} &a_{6} &a_{7} &a_{8} \\
\end{pmatrix}
\begin{pmatrix}
\delta \phi \\
\delta \chi \\
\delta x \\
\delta y \\
\end{pmatrix} \, .
\ee
Here
\begin{align}
\label{two16}
a_{1} =& \frac{2HH'-2H'^2+3m^3\omega_{,\phi}H^3H'
+ m^2\kappa^2H^2(1+m^2\omega H^2)}{H(1+3m^2\omega H^2)} \nn
& + \frac{1}{H}\left( 4HH'+3H'^2+HH''
+\frac{m^2\kappa^2}{2}(1+m^2\omega H^2)(H^2+HH')\right) \, ,\nn
a_{2} =& \frac{m^2\kappa^2(1+m^2\omega H^2)}{H(1+3m^2\omega H^2)}
\left( \frac{1}{2}(H^2+HH')(-1+3m^2\eta H^2)-H^2\right) \, ,\nn
a_{3} =& -3 - \frac{m^2\kappa^2}{2}(1+m^2\omega H^2)
 - \frac{6m^2\omega H^2H'+3m^3\omega_{,\phi}H^3}{H(1+3m^2\omega H^2)} \, ,\nn
a_{4} =& -\frac{m^2\kappa^2}{2}\frac{1+m^2\omega H^2}{1+3m^2\omega H^2}(-1+3m^2\eta H^2) \, ,\nn
a_{5} =& \frac{m^2\kappa^2(-1+m^2\eta H^2)}{H(-1+3m^2\eta H^2)}
\left( \frac{1}{2}(H^2+HH')(1+3m^2\omega H^2)+H^2\right) \, ,\nn
a_{6} =& -\frac{2HH'-2H'^2-3m^3\eta_{,\chi}H^3H'
+m^2\kappa^2H^2(-1+m^2\eta H^2)}{H(-1+3m^2\eta H^2)} \nn
& + \frac{1}{H}\left( 4HH'+3H'^2+HH''
+\frac{m^2\kappa^2}{2}(-1+m^2\eta H^2)(H^2+HH')\right) \, ,\nn
a_{7} =& -\frac{m^2\kappa^2}{2}\frac{-1+m^2\eta H^2}{-1+3m^2\eta H^2}(1+3m^2\omega H^2) \, ,\nn
a_{8} =& -3 - \frac{m^2\kappa^2}{2}(-1+m^2\eta H^2)
 - \frac{6m^2\eta H^2H'+3m^3\eta_{,\chi}H^3}{H(-1+3m^2\eta H^2)} \, ,
\end{align}
where we have used the constraint equation between the perturbations, 
which is obtained from (\ref{two4}) as follows, 
\bea
\label{two17}
\delta H &=& \frac{\kappa^2\delta \rho_s}{6H} \nn
&=& \dot \phi(1+3\omega \dot \phi^2)\delta x 
+ \left( \frac{3}{4}\omega_{,\phi}\dot \phi^4+V_{,\phi}\right)\delta \phi 
+ \dot \chi (-1+3\eta \dot \chi^2)\delta y + \left( \frac{3}{4}\eta_{,\chi}\dot 
\chi^4+V_{,\chi}\right)\delta \chi\, .
\eea
In order the solution  (\ref{two9}) is stable, the real parts of all the eigenvalues of 
the above matrix (\ref{two15}) must be negative. 
Then we find the conditions that the real parts of all the eigenvalues are negative. 
The characteristic equation and its solutions are given as follows:
\bea
\label{two18}
&\lambda^4& +A_{1}\lambda^3 + A_{2}\lambda^2 + A_{3}\lambda +A_{4} = 0, \nn 
&\lambda& = \frac{1}{2}\left( +q\pm \sqrt{q^2-4\Xi \pm 4p}\right) - \frac{A_{1}}{4},\ 
                 \frac{1}{2}\left( -q\pm \sqrt{q^2-4\Xi \pm 4p}\right) - \frac{A_{1}}{4} .
\eea
Here
\bea
\label{two19}
&& A_{1} = -a_{3}-a_{8} \, , \quad 
A_{2} = a_{3}a_{8}-a_{4}a_{7}-\frac{a_{1}+a_{6}}{H} \, ,\nn
&& A_{3} = \frac{1}{H}(a_{1}a_{8}+a_{3}a_{6}-a_{4}a_{5}-a_{2}a_{7}) \, ,\quad
A_{4} = \frac{1}{H^2}(a_{1}a_{6}-a_{2}a_{5}) \, .
\eea 
Here $\Xi$ is one of the solution of the following cubic equation;
\be
\label{two20}
\Xi^3 -\frac{B_{1}}{2}\Xi^2 -B_{3}\Xi +\frac{B_{1}B_{3}}{2}-\frac{1}{8}B_{2}^2 = 0 \, ,
\ee
that is
\be
\label{two21}
\Xi = \left( -P-\sqrt{P^2+Q^3}\right)^{\frac{1}{3}} + \left( -P+\sqrt{P^2+Q^3}\right)^{\frac{1}{3}} 
+\frac{B_{1}}{6} \, .
\ee
Here
\bea
\label{two22}
&& B_{1} =  -\frac{3A_{1}^2}{8}+A_{2} \, , \quad 
B_{2} =  \frac{A_{1}^3}{8}-\frac{A_{1}A_{2}}{2}+A_{3} \, , \quad 
B_{3} =  -\frac{3A_{1}^4}{256}+\frac{A_{1}^2A_{2}}{16}-\frac{A_{1}A_{3}}{4}+A_{4} \nn
&& P =  -\frac{B_{1}^3}{216}+\frac{B_{1}B_{3}}{6}-\frac{B_{2}^2}{16} \, , \quad 
Q =  -\frac{B_{1}^2}{36} -\frac{B_{2}}{3} \, .
\eea
In (\ref{two18}), ($p,\, q $) are defined by
\bea
\label{two23}
p = \sqrt{\Xi-B_{3}} \, , \quad q =  \sqrt{2\Xi-B_{1}}\, .
\eea
Now we consider the real solution of the cubic equation (\ref{two20}) since we like to find 
the condition for the maximum value of the real parts of $\lambda$ 
(denoted as $\Re \lambda_\mathrm{max}$) to be negative.
Then, all the quantities given by the capital characters in (\ref{two19}) and (\ref{two22}) 
become real numbers. Then we find the expression of $\Xi $ for four cases, 
\begin{align}
\label{two24}
\Xi = \begin{cases}
2|Q|^{\frac{1}{2}}\mathrm{cos}\left [ \frac{1}{3}
\mathrm{arccos}\left(\frac{-P}{|Q|^\frac{3}{2}}\right)\right] + \frac{B_{1}}{6}
& \mathrm{for}\ Q<0\ \mathrm{and}\ P^2<|Q^3| \\
 -\left( P+\sqrt{P^2+Q^3}\right)^{\frac{1}{3}} - \left( P-\sqrt{P^2+Q^3}\right)^{\frac{1}{3}} +\frac{B_{1}}{6}
& \mathrm{for}\ Q<0,\ P^2>|Q^3|\ \mathrm{and}\ P\geq 0 \nn
\left( -P+\sqrt{P^2+Q^3}\right)^{\frac{1}{3}} + \left( -P-\sqrt{P^2+Q^3}\right)^{\frac{1}{3}} +\frac{B_{1}}{6}
& \mathrm{for}\ Q<0,\ P^2>|Q^3|\ \mathrm{and}\ P<0 \\
 -\left( P+\sqrt{P^2+Q^3}\right)^{\frac{1}{3}} + \left( -P+\sqrt{P^2+Q^3}\right)^{\frac{1}{3}} +\frac{B_{1}}{6} 
& \mathrm{for}\ Q>0 \, .
\end{cases}
\end{align}
We also find the expression of $\Re \lambda_\mathrm{max}$ as follows, 
\begin{align}
%\label{two25}
\intertext{For $\Xi^2\geq B_{3}$,}
\Re\lambda_\mathrm{max} &= 
\begin{cases}
\frac{1}{2}\left( q+\sqrt{-2\Xi -B_{1}+4p}\right) -\frac{A_{1}}{4}
& \mathrm{for}\ \Xi\geq \frac{B_{1}}{2}\ \mathrm{and}\ -2\Xi -B_{1}+4l \geq 0\nn
\frac{1}{2}q -\frac{A_{1}}{4} \leq 0
& \mathrm{for}\ \Xi\geq \frac{B_{1}}{2}\ \mathrm{and}\ -2\Xi -B_{1}+4l < 0 \nn
\frac{1}{2}\sqrt{-2\Xi -B_{1}+4p} -\frac{A_{1}}{4} 
& \mathrm{for}\ \Xi < \frac{B_{1}}{2}\ \mathrm{and}\ -2\Xi -B_{1}+4l\geq 0 \nn
 -\frac{A_{1}}{4} 
& \mathrm{for}\ \Xi < \frac{B_{1}}{2}\ \mathrm{and}\ -2\Xi -B_{1}+4l < 0
\end{cases} \nn
&\leq 0\, , \nn
\intertext{for $\Xi^2<B_{3}$,}
\Re\lambda_\mathrm{max} &= 
\begin{cases}
\frac{1}{2}\left( q+\sqrt{\frac{1}{2}
\left(-2\Xi -B_{1}+\sqrt{(2\Xi+B_{1})^2+16p^2}\right)}\right) -\frac{A_{1}}{4}
& \mathrm{for}\ \Xi\geq \frac{B_{1}}{2} \nn
\frac{1}{2}\sqrt{\frac{1}{2}\left(-2\Xi -B_{1}+\sqrt{(2\Xi+B_{1})^2+16p^2}\right)} 
 -\frac{A_{1}}{4}
& \mathrm{for}\ \Xi < \frac{B_{1}}{2}
\end{cases} \nn
&\leq 0\, .
\end{align}
This condition may reduce to the constraints on $\alpha'(mN)$, $\alpha(mN)$, $f(N)$ in $\omega_{,\phi}(mN)$ etc.
However, it is not straightforward to obtain the constraints on $\alpha'(mN)$ for each of models (or equivalently, 
for each of $f(N)$) explicitly because the condition, $P^2>|Q^3|$, is the 12th order inequality with respect to 
$\alpha'(mN)$ and we cannot solve it analytically in general. 
Furthermore, as in the case of the one scalar k-essence model \cite{Saitou:2011hv}, 
it is very difficult to find the explicit form of 
$\alpha(mN)$ even if we obtain the explicit constraints of $\alpha'(mN)$
since it is not an equality but an inequality.
Therefore, from the condition for the stability of the solution, 
it is difficult to find the explicit form of the function $\alpha(mN)$ for each of models.
We can obtain, however, more explicit constraint for $\alpha(mN)$ from the condition that the square of 
the sound speed should be positive, as we will see in the next subsection. 

\subsection{Constraints from sound speed}

We now consider the stability constraints coming from the sound speed of the scalar fields. 
By considering the perturbation of the scalar fields, we find the expressions of 
the sound speed $c_{si}^2$ ($i=\phi,\, \chi$) as follows,
\be
\label{two26}
c_{s\phi}^2 = \frac{p_{\phi,X}}{\rho _{\phi,X}}= \frac{1+2\omega X}{1+6\omega X} \, ,\quad 
c_{s\chi}^2 = \frac{p_{\chi,U}}{\rho _{\chi,U}}= \frac{-1+2\eta U}{-1+6\eta U} \, .
\ee
The sound speed must be $0\leq c_{si}^2 (\leq 1)$ for the stability. 
This condition can be expressed as the constraints for the function $\alpha$ 
for the reconstructed solution (\ref{two9}),
\bea
\label{two27}
&& \alpha(\phi )|_{\phi = mN} \geq \frac{f'}{3m^4H^4} \quad \mbox{or} \quad
\alpha(\phi )|_{\phi = mN} \leq \frac{f'}{3m^4H^4} - \frac{1}{m^2H^2} \, ,\nn
\mbox{and} && \alpha(\chi )|_{\chi = mN} \geq 0 \quad \mbox{or} \quad
\alpha(\chi )|_{\chi = mN} \leq -\frac{1}{m^2H^2} \, .
\eea

Although the models in this paper could be regarded as an effective and classical 
low energy theory of more fundamental theories, we may also consider the (in)stability 
as a quantum theory. 
As a quantum theory, we require the conditions $p_{\phi,X}>0$, $\rho _{\phi,X}>0$,
$p_{\chi,U}>0$, and $\rho _{\chi,U}>0$ 
for stability and the conditions give the constraints $\alpha(mN) \leq -\frac{1}{m^2H^2}$ 
and $\alpha(mN) \geq \frac{f'}{3m^4H^4}$.
If the universe 
has the phantom crossing time, however, the derivative of the energy density for 
the scalar fields $f'$ in (\ref{two9}) changes its sign from negative to positive 
and therefore the reconstructed solution (\ref{two9}) 
of the evolution of the universe cannot be stable as a quantum theory at the phantom crossing time. 
We should note, however, that this analysis of the sound speed is valid only 
in the sub-horizon scale and in absence of the gravity because we do
not consider the FLRW space-time and the gravitational fluctuations. 
Therefore, we do not take care of the quantum instability here and we only consider the 
classical stability of the models in the next section.

\section{Two scalar models \label{SecV}}

\subsection{Model 1}

As a first example of the model with phantom crossing, we consider the following model 
with matters:
\bea
\label{two28}
&H^2(N)& =  H_0^2\left( \Omega_{\mathrm{DE}\,0}\frac{\cosh[\gamma(N-N_c)]}{\cosh[\gamma(N_0-N_c)]}
+ \Omega_{m0}\e^{-3(N-N_0)}\right)\, , \\
&\Omega_{DE0}& = 0.74\, ,\quad \Omega_{m0} = 0.26\, ,
\eea
where subscript ``$0$'' denotes the present value of each parameter and the free parameter $N_c$ is
the e-folding number at the phantom crossing time. 
The parameter $\gamma$ is to be determined by the observational data and the stability constraints. 
The model (\ref{two28}) reduces to $\Lambda$CDM model in the limit of $\gamma = 0$.

At the point of the phantom crossing $N=N_0$, we find 
\be
\label{TwoTwo1}
H=H_0\, \quad \dot H = 0\, ,\quad 
\ddot H = \frac{H_0^3 \Omega_{\mathrm{DE}\,0} \gamma^2}
{2 \cosh[\gamma(N_0-N_c)]}\, .
\ee
Therefore if we try to realize the development of the universe given by (\ref{two28}) by 
using one scalar model (\ref{ma7}), the eigenvalue $M_+$ in (\ref{R6}) diverges 
at the crossing point and therefore the infinite instability is generated there. 
This tells that the development in (\ref{two28}) cannot be realized by one scalar model. 
In case of two scalar model, however, as we will see soon, the reconstructed solution becomes 
stable and the universe can develop as in (\ref{two28}). 

In order to find the viable region of the parameter $\gamma$, we introduce cosmological parameters, 
the equation of state parameter $w_\mathrm{DE}$, the deceleration parameter $q$, the jerk 
parameter $j$, and the snap parameter $s$, whose present values are constrained 
by observations \cite{Amanullah:2010vv}:
\be
\label{two29}
w_\mathrm{DE} = -1 -\frac{f'}{3f} \, , \quad
q \equiv -\frac{1}{a}\ddot{a}H^{-2} = -1+\frac{3}{2}\Omega_{m0} -\frac{\kappa^2f'}{6H_0^2} \, ,\quad
j \equiv  \frac{1}{a}\frac{d^3a}{dt^3}H^{-3} \, ,\quad
s \equiv \frac{1}{a}\frac{d^4a}{dt^4}H^{-4}\, .
\ee  
The observational constraints to these parameters are shown in Table \ref{Table1}.
\begin{table}
\centering 
\caption{The constraints for the cosmological parameters from the 
cosmological observation \cite{Amanullah:2010vv}.}\label{Table1}
\begin{tabular}{c|c}
\hline
% after \\ : \hline or \cline{col1-col2} \cline{col3-col4} ...
observables & 68\% CL   \\
\hline
$w_{\mathrm{DE}\,0}$ & $-1.032\pm0.144$   \\
$q_{0}$ & $(-0.60,\ -0.30)$ \\
$j_{0}$  & $(-0.88,\ 0.90)$ \\
$s_{0}$ & $(-0.57,\ 1.07)$ \\
\hline
\end{tabular}
\end{table}
In case of our model, the observable parameters can be expressed as
\begin{align}
\label{two30}
w_{\mathrm{DE}\,0} &= -1 -\frac{\gamma}{3}\tanh[\gamma(N_0-N_c)]\, , \nn
q_0 &= -1+\frac{3}{2}\Omega_{m0} 
 -\frac{\gamma}{2}\Omega_{\mathrm{DE}\,0}\tanh[\gamma(N_0-N_c)] \, ,\nn
j_0 &= 1+ \Omega_{\mathrm{DE}\,0}\left(\frac{1}{2}\gamma 
+3\gamma\tanh[\gamma(N_0-N_c)]\right) \, ,\nn
s_0 &= \frac{1}{2}\left(\Omega_{\mathrm{DE}\,0}\gamma^3\tanh[\gamma(N_0-N_c)]-27\Omega_{m0}\right)
+(j_0+3q_0+2)(3-q_0)+3q_0^2-2 \, .
\end{align}
%Here the suffix ``$0$'' expresses the values in the present universe. 
We show the viable region of the parameter $\gamma$ for various values of $N_c$ 
in Table \ref{Table2}.
\begin{table}
\centering 
\caption{The viable region of the parameter $\gamma$ for various values of $N_c$.}
\label{Table2}
\begin{tabular}{c|c|c}
\hline
$N_c-N_0$ & $t_c - t_0$ & $|\gamma|$  \\
\hline
$1.0$ & $\sim 25.2$\,Gyrs & $0.236\leq|\gamma|\leq 0.306$  \\
$5.0$ & $\sim 150.2$\,Gyrs & $0.101\leq|\gamma|\leq 0.129$  \\
$10$ & $\sim 300.7$\,Gyrs & $0.074\leq|\gamma|\leq 0.096$  \\
\hline
\end{tabular}
\end{table}
In order that the observational constraints could be satisfied, 
the phantom crossing must occur {\it in future} for any model (not in {\it past}), 
which is clear from the form of parameter $q$ in (\ref{two29}). 
When we change the variable form the cosmic time $t$ to the e-folding number $N$, 
we used the Hubble parameter in $\Lambda$CDM model as an approximation  
since $\gamma $ is nearly $0$ for each case and 
therefore the energy density of the scalar fields do not vary so fast.
Then the relation between the cosmic time $t$ and the e-folding number $N$ is now 
given by
\be
\label{two31}
H(t) \simeq \frac{2}{3}\sigma\coth[\sigma t] \, ,\quad
N_0-N = \int^{t_0}_{t}H(t)dt 
\simeq \frac{2}{3}\ln\frac{\sinh[\sigma t_0]}{\sinh[\sigma t]}\, ,
\ee
where $\sigma = \frac{\kappa}{2}\sqrt{3\Lambda}$.
We assume the universe was dominated by the matter from the time $t_m$.
When $t_0 = 13.7\times 10^9$\,yrs and $t_m = 7\times 10^4$\,yrs, 
we obtain $N_0-N_m \simeq 8.16$.
If we assume that the expression in (\ref{two31}) 
is valid when $N_c-N_0\lesssim 10$,
we can estimate the crossing time $t_c$ for each $N_c-N_0$ as shown 
in Table \ref{Table2}.

\begin{figure}[h]
\begin{center}
\includegraphics[width=4in]{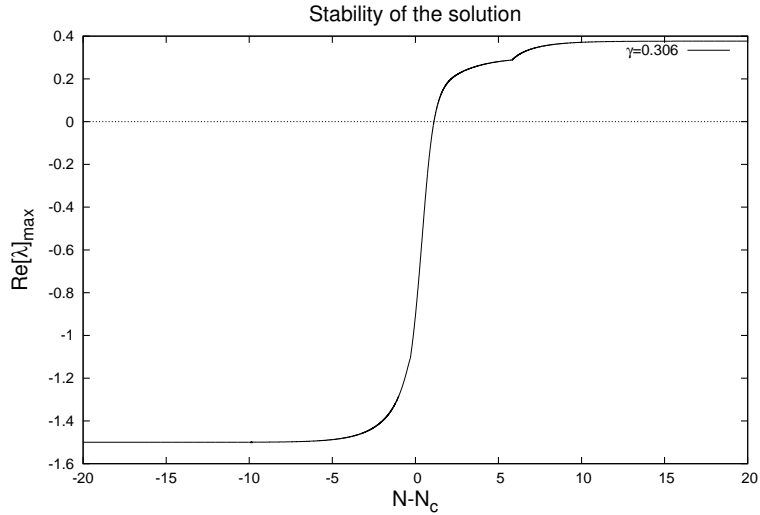}
\caption{$\Re\lambda_{max}-(N-N_c)$ curve in the case of $\alpha(mN)=\frac{1}{m^2H^2}$
and $\gamma = 0.306$. The region $\Re\lambda_{max}<0$ is stable but 
$\Re\lambda_{max}>0$ is unstable. This graph shows model 1 is stable for a few e-folding number
after phantom crossing when $N-N_c = 0$.}
\label{Figure1}
\end{center}
\end{figure}

As a second step, we try to find the function $\alpha(mN)$ so that the 
reconstructed solution (\ref{two9}) could be stable. 
We investigate two cases, $\alpha(mN) = -\frac{2}{m^2H^2}$ and 
$\alpha(mN) = +\frac{1}{m^2H^2}$ by putting $m = \frac{1}{\kappa}$. 
For both of the cases, we find that the solution is stable 
for ten or hundreds billion years (depending on the value of $N_c$) after the 
phantom crossing but becomes unstable at a later time. See Figure \ref{Figure1} 
and Table \ref{Table4}.

\begin{table}[h]
\centering 
\caption{The limit values of $N-N_c$ and $t-t_c$ when model 1 is stable
in the case $\alpha(mN) = \frac{1}{m^2H^2}$. We obtain similar result also in the case 
$\alpha(mN) = -\frac{2}{m^2H^2}$.}
\label{Table4}
\begin{tabular}{|c|c|c|c|}
\hline
$N_c-N_0$ & \multicolumn{2}{|c|}{Stable limit } & $|\gamma|$  \\
\hline
$1.0$ & $N-N_c = +^{1.253}_{1.12}$& $t-t_c \sim +^{38.8}_{34.6}$\,Gyrs & $|\gamma|=^{0.236}_{0.306}$  \\
\hline
$5.0$ & $N-N_c = +^{5.524}_{5.402}$& $t-t_c \sim +^{173.2}_{169.4}$\,Gyrs & $|\gamma|=^{0.101}_{0.129}$  \\
\hline
$10$ & $N-N_c = +^{10.569}_{10.440}$& $t-t_c \sim +^{331.4}_{327.3}$\,Gyrs & $|\gamma|=^{0.074}_{0.096}$  \\
\hline

\end{tabular}
\end{table}
Finally, we explicitly give the reconstructed Lagrangian of the scalar fields which generates 
the stable phantom crossing
in the case $m = \frac{1}{\kappa}$ and $\alpha = \frac{A}{m^2H^2}$:
\begin{align}
\mathcal{L}_s =& X + \left(-\frac{\kappa^2 \gamma H_0^2\Omega_{DE0}}{\cosh[\kappa \gamma(\phi_0-\phi_c)]}
\frac{\sinh[\kappa \gamma(\phi-\phi_c)]}{H^4(\kappa \phi)}+ \frac{\kappa^2A}{H^2(\kappa \phi)}\right)X^2 
-U -\frac{\kappa^2A}{H^2(\kappa \chi)}U^2 \nn
&- \frac{1}{\kappa^2}\left( \frac{H^2_0\Omega_{DE0}}
{\cosh[\kappa \gamma(\phi_0-\phi_c)]}\right) \left \{ \left( 3-\frac{3A+2}{4}\right)
\cosh[\kappa \gamma(\phi-\phi_c)]+\left( \frac{3\gamma}{4}-\frac{3A+3}{\gamma}\right)
\sinh[\kappa \gamma(\phi-\phi_c)]\right \} \nn
&- \frac{H_0^2\Omega_{m0}(A+2)}{4\kappa^2}\e^{-3\kappa(\phi-\phi_0)} \nn
&- \frac{1}{\kappa^2}\left( \frac{H^2_0\Omega_{DE0}} 
{\cosh[\kappa \gamma(\chi_0-\chi_c)]}\right) \left \{ \frac{3A+2}{4}\cosh[\kappa \gamma(\chi-\chi_c)]+
\frac{3A+3}{\gamma}\sinh[\kappa \gamma(\chi-\chi_c)]\right \} \nn
&+ \frac{H_0^2\Omega_{m0}(A+2)}{4\kappa^2}\e^{-3\kappa(\chi-\chi_0)}\, ,
\end{align}
where $A = 1,\ -2$.

\subsection{Model 2}

As a second model, we consider the following model which may have the Little Rip 
\cite{Frampton:2011sp,Frampton:2011rh,Nojiri:2011kd,arXiv:1107.4642,Astashenok:2012tv,
arXiv:1103.2480,arXiv:1111.2454,Frampton:2011aa}:
\be
\label{LRm}
H^2(N) = H_0^2[\Omega_{DE0}\frac{\e^{-\delta (N-N_c)}+\delta(N-N_c)} 
{\e^{N_0-N_c}+\delta(N_0-N_c)}+\Omega_{m0}\e^{-3(N-N_0)}]
\ee
where $\delta $ is the parameter which should be constrained.
The obtained constraints are given in Table \ref{Table3}.
\begin{table}
\centering 
\caption{The viable region of the parameter $\delta $ for the model 2 and 
$\epsilon$ for the model 3 for various values of $N_c$.}
\label{Table3}
\begin{tabular}{c|c|c}
\hline
$N_c-N_0$ & $t_c - t_0$ & $\delta \ \mathrm{or}\  \epsilon $  \\
\hline
$1.0$ & $\sim 25.2$\,Gyrs & $0.37\leq \delta \ \mathrm{or} \ \epsilon \leq 0.89$  \\
$5.0$ & $\sim$\,150.2Gyrs & $0.14\leq \delta \ \mathrm{or} \ \epsilon \leq 0.28$  \\
$10$ & $\sim 300.7$\,Gyrs & $0.10\leq \delta \ \mathrm{or} \ \epsilon \leq 0.26$  \\
\hline
\end{tabular}
\end{table}
Then, as in model 1, we choose $\alpha(mN) = \frac{A}{m^2H^2(N)}$ (where $A=1,\,-2$) and we obtain 
results similar to those for model 1 about the stability, that is, the solution becomes 
unstable in ten or hundreds billion years
after phantom crossing. This means that the universe may evolves to other future
without the Little Rip singularity for $\alpha(mN) = \frac{A}{m^2H^2(N)}$.
The reconstructed Lagrangian of the scalar fields
in the case $m = \frac{1}{\kappa}$ and $\alpha = \frac{A}{m^2H^2}$ is given as follows:
\begin{align}
\mathcal{L}_s &= X + \left(\frac{\kappa^2 \delta H_0^2\Omega_{DE0}}
{\e^{-\kappa \delta(\phi_0-\phi_c)}+\kappa \delta(\phi_0-\phi_c)}
\frac{\e^{-\kappa \delta (\phi-\phi_c)}-1}{H^4(\kappa \phi)}+ \frac{\kappa^2A}{H^2(\kappa \phi)}\right)X^2 
-U -\frac{\kappa^2A}{H^2(\kappa \chi)}U^2 \nn
&- \frac{1}{\kappa^2}\left( \frac{H^2_0\Omega_{DE0}}
{\e^{-\kappa \delta(\phi_0-\phi_c)}+\kappa \delta(\phi_0-\phi_c)}\right) 
\bigg\{ \left( 3-\frac{3A+2}{4}+\frac{3A+3}{\delta}-\frac{3\delta}{4}\right)
\e^{-\kappa \delta (\phi-\phi_c)}-\frac{3\kappa^2\delta(A+1)}{2}(\phi-\phi_c)^2 \nn
&+\kappa \delta \left( 3-\frac{3A+2}{4}\right)(\phi-\phi_c)+\frac{3\delta}{4} \bigg\}
- \frac{H_0^2\Omega_{m0}(A+2)}{4\kappa^2}\e^{-3\kappa(\phi-\phi_0)} \nn
&- \frac{1}{\kappa^2}\left( \frac{H^2_0\Omega_{DE0}} 
{\e^{-\kappa \delta(\chi_0-\chi_c)}+\kappa \delta(\chi_0-\chi_c)}\right) 
\bigg \{ \left(\frac{3A+2}{4}-\frac{3A+3}{\delta }\right)\e^{-\kappa \delta (\chi-\chi_c)}+
\frac{3\kappa^2 \delta(A+1)}{2}(\chi-\chi_c)^2 \nn
&+\frac{\kappa \delta(3A+2)}{4}(\chi-\chi_c) \bigg \}
+ \frac{H_0^2\Omega_{m0}(A+2)}{4\kappa^2}\e^{-3\kappa(\chi-\chi_0)}\, .
\end{align}

\subsection{Model 3}

As a third model, we suggest the model which may have asymptotically de Sitter space.
The model has the following form:
\be
H^2(N) = H_0^2[\Omega_{DE0}\frac{\e^{-\epsilon (N-N_c)}+\frac{\epsilon}{\theta}
(1-\e^{-\theta (N-N_c)})} 
{\e^{-\epsilon (N_0-N_c)}+\frac{\epsilon}{\theta}
(1-\e^{-\theta (N_0-N_c)})}+\Omega_{m0}\e^{-3(N-N_0)}]
\ee
where $\epsilon $ and $\theta (< \epsilon)$ are the parameters which should be constrained.
We set $\theta = 0.01$ for simplicity. In this case, we obtain the same result as 
in the Little Rip model (\ref{LRm}), which is shown in Table \ref{Table3}.
As former models, we choose $\alpha(mN) = \frac{A}{m^2H^2(N)}$ (where $A=1,\,-2$) and we obtain the result 
similar to former models (\ref{two28}) and (\ref{LRm}) about the stability, that is, 
the solution becomes unstable in ten or hundreds billion years after phantom crossing. 
This tells that the universe may evolves not to asymptotically de Sitter space
for $\alpha(mN) = \frac{A}{m^2H^2(N)}$.
The reconstructed Lagrangian of the scalar fields
in the case $m = \frac{1}{\kappa}$ and $\alpha = \frac{A}{m^2H^2}$ is given as follows:
\begin{align}
\mathcal{L}_s &= X + \left(\frac{\kappa^2 \epsilon H_0^2\Omega_{DE0}}
{\e^{-\kappa \epsilon(\phi_0-\phi_c)}+\frac{\epsilon}{\theta}(1-\e^{-\kappa \epsilon(\phi_0-\phi_c)})}
\frac{\e^{-\kappa \epsilon (\phi-\phi_c)}-\e^{-\kappa \theta (\phi-\phi_c)}}
{H^4(\kappa \phi)}+ \frac{\kappa^2A}{H^2(\kappa \phi)}\right)X^2 
-U -\frac{\kappa^2A}{H^2(\kappa \chi)}U^2 \nn
&- \frac{1}{\kappa^2}\left( \frac{H^2_0\Omega_{DE0}}
{\e^{-\kappa \epsilon(\phi_0-\phi_c)}+\frac{\epsilon}{\theta}(1-\e^{-\kappa \epsilon(\phi_0-\phi_c)})}\right) 
\Bigg[ \left( 3-\frac{3A+2}{4}+\frac{3A+3}{\epsilon}-\frac{3\epsilon}{4}\right)
\e^{-\kappa \epsilon (\phi-\phi_c)} \nn
&- \frac{\epsilon}{\theta}\bigg\{\left( 3-\frac{3A+2}{4}+\frac{3A+3}{\theta}-\frac{3\theta}{4}\right)
\e^{\kappa \theta (\phi-\phi_c)} + (3A+3)\kappa \phi -3+\frac{3A+2}{4}\bigg\}\Bigg]
- \frac{H_0^2\Omega_{m0}(A+2)}{4\kappa^2}\e^{-3\kappa(\phi-\phi_0)} \nn
&- \frac{1}{\kappa^2}\left( \frac{H^2_0\Omega_{DE0}} 
{\e^{-\kappa \epsilon(\chi_0-\chi_c)}+\frac{\epsilon}{\theta}(1-\e^{-\kappa \epsilon(\chi_0-\chi_c)})}\right) 
\Bigg [ \left(\frac{3A+2}{4}-\frac{3A+3}{\epsilon }\right)\e^{-\kappa \epsilon (\chi-\chi_c)} \nn
&+ \frac{\epsilon}{\theta}\bigg\{\left( \frac{3A+3}{\theta}-\frac{3A+2}{4}\right)
\e^{\kappa \theta (\chi-\chi_c)} + (3A+3)\kappa \chi +\frac{3A+2}{4}\bigg\}\Bigg]
+ \frac{H_0^2\Omega_{m0}(A+2)}{4\kappa^2}\e^{-3\kappa(\chi-\chi_0)}\, .
\end{align}

\section{Summary}

In this paper, we considered the models with two scalar fields, 
which has the structure as in the ghost condensation model 
\cite{ArkaniHamed:2003uy,ArkaniHamed:2003uz} or $k$-essence 
model \cite{Chiba:1999ka,ArmendarizPicon:2000dh,ArmendarizPicon:2000ah}. 
The models can describe the stable phantom crossing, which should be 
contrasted with one scalar tensor models, where the large instability 
occurs at the crossing the phantom divide. 
In the previous two scalar models, which are extensions of the one 
scalar tensor model, it was difficult to give a general 
formulation of the reconstruction when we include 
{\it matters}, which is realized in this paper in terms of the e-foldings $N$. 
We also gave general arguments for the stabilities of the models and the 
reconstructed solution. We also investigate the viability of the models by 
comparing the observational data. 

\section*{Acknowledgments}

We are grateful to S.~D.~Odintsov for the discussion when he stayed in 
Nagoya University. 
This research has been supported in part
by Global COE Program of Nagoya University (G07)
provided by the Ministry of Education, Culture, Sports, Science \&
Technology and by the JSPS Grant-in-Aid for Scientific Research (S) \# 22224003
and (C) \# 23540296 (SN).

\end{document}